\documentclass[conference]{IEEEtran}
\IEEEoverridecommandlockouts
\usepackage{amsmath,amssymb,amsfonts}
\usepackage{algorithm}
\usepackage{algorithmic}
\usepackage{makecell}
\usepackage{graphicx}
\usepackage{tabularx}
\usepackage{threeparttable}
\usepackage{xcolor}
\usepackage{pifont}
\definecolor{mygreen}{RGB}{146,208,80}
\definecolor{myred}{RGB}{192,0,0}
\newcommand{\cmark}{\textcolor{mygreen}{\ding{51}}}
\newcommand{\xmark}{\textcolor{myred}{\ding{55}}}

\usepackage{booktabs}
\usepackage{verbatim}
\usepackage{textcomp}
\usepackage{multirow}

\usepackage{cite}
\def\BibTeX{{\rm B\kern-.05em{\sc i\kern-.025em b}\kern-.08em
    T\kern-.1667em\lower.7ex\hbox{E}\kern-.125emX}}
\begin{document}

\title{BenDi: An Energy-Efficient Quasi-Stochastic Systolic Architecture for Edge Bioelectronics}

\author{\IEEEauthorblockN{Bochen Ye, Yihan Pan, Shady Agwa, Themis Prodromakis}
}
\linespread{0.85}
\maketitle
 \begin{abstract}
Continuous long-term monitoring and diagnosis of biomedical signals, such as electrocardiograms (ECGs), can help mitigate an increasing threat to public health. Artificial Intelligence (AI) models, such as Convolutional Neural Networks (CNNs), provide accurate monitoring and classification for relevant diseases; however, they require more computational resources than conventional AI hardware can typically afford, especially for a resource-constrained environment on the edge. 
In this work, we present BenDi, an energy-efficient quasi-stochastic systolic architecture for bioelectronic systems on the edge. BenDi leverages multiple levels of energy and power optimization, ranging from circuits to software quantization, including low supply voltage, the \underline{Ben}t-Pyramid data format for quasi-stochastic multiplication, the \underline{Di}P systolic dataflow, and hardware-aware quantization, to handle CNNs with high accuracy on the edge within limited hardware budgets.
The hardware implementation results, using a commercial 22nm technology, show that BenDi architecture, at 0.5 Voltage and 100 MHz, offers 3.35x smaller area and 5x higher energy efficiency, compared to state-of-the-art binary-based weight-stationary systolic architectures.
Regarding Bioelectronic edge systems, BenDi achieves an order-of-magnitude improvement in energy efficiency and another order-of-magnitude improvement in area efficiency, compared to its counterparts. This significant improvement comes at the cost of 1\% to 3.3\% accuracy loss on the MIT-BIH and Apnea-ECG benchmarks, respectively, compared with conventional computing using the 32-bit floating-point format.

\end{abstract}

\begin{IEEEkeywords}
Quasi-stochastic Computing, Systolic Array, Bent-Pyramid, Bioelectronics, AI, CNN.
\end{IEEEkeywords}

\section{Introduction}
\label{sec:intro}

Growing public health challenges have made continuous long-term monitoring and diagnosis of biomedical signals, including electrocardiograms (ECGs), increasingly important.
As a representative physiological signal, the ECGs provides rich diagnostic information for a wide range of clinical applications, including arrhythmia detection and sleep apnea assessment.  
Artificial Intelligence (AI), particularly Convolutional Neural Networks (CNNs), has demonstrated strong potential for biomedical signal processing through automatic feature extraction and classification~\cite{ecg}. However, deploying these CNN models on resource-constrained edge devices remains a significant challenge due to strict constraints on power, energy, and silicon area.

One major challenge is that CNN models incorporate millions of parameters and are predominantly characterized by Multiply-Accumulate (MAC) operations. 
In Table~\ref{tab:paradigm_comparison}, multiplication operations of binary computing (BC) are executed deterministically, maintaining precision, therefore ensuring the accuracy of the final output.
The state-of-the-art hardware implementation usually employs systolic array architectures to perform matrix multiplication using a flexible tiling strategy. 
However, MAC units within systolic array are complex and power-hungry components for binary arithmetic, such as floating-point and integer formats. These MAC units are resource-intensive, demanding higher power, energy, and area budget on the edge.

\begin{table}[!t]
\centering
\caption{Comparison of Different Computing Paradigms}
\label{tab:paradigm_comparison}
\begin{tabular}{lccc}
\toprule
Paradigm & Determinism & Hardware Cost & Latency \\
\midrule
Binary        & Deterministic \cmark & High \xmark & Low \cmark \\
Stochastic    & Non-deterministic \xmark & Very Low \cmark & High \xmark \\
\textbf{Our Work} &\textbf{Deterministic \cmark}  & \textbf{Low \cmark} & \textbf{Low \cmark} \\
\bottomrule
\end{tabular}
\end{table}

\begin{figure}[!t]
    \centering
    \includegraphics[width=0.8\linewidth]{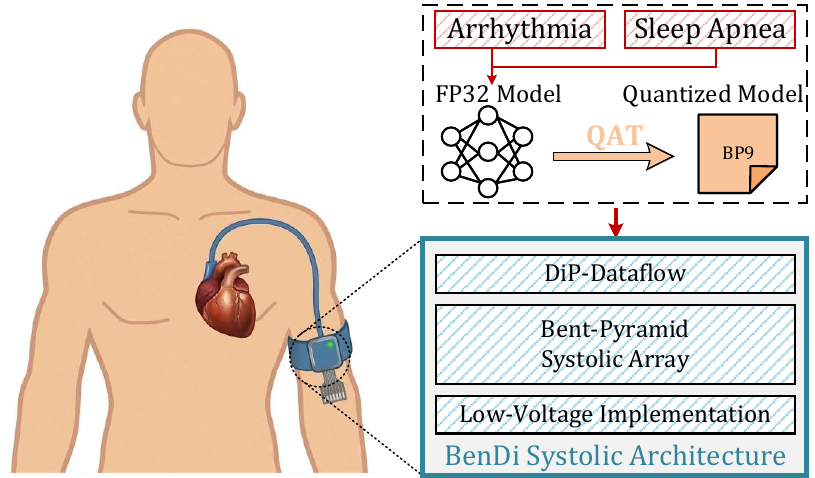}
    \caption{Overview of hardware-software co-design flow. The software will train model in FP32 and then quantization-aware training with BP9 data format. The proposed BenDi architecture can support arrhythmia detections and sleep apnea detections.}
    \label{fig:f1}
\end{figure}

Stochastic computing (SC) is proposed as a potential solution to reduce the computational complexity by representing data as a bitstream of 1s and 0s that is equivalent to a probability~\cite{tvlsi25}. 
SC offers significant improvement in energy and area efficiency over BC through simple AND gates for unipolar SC numbers, at the cost of sacrificing deterministic computing and accuracy.
Moreover, traditional SC hardware incorporates an expensive hardware overhead due to SC Number Generators (SNGs) and incurs high latencies because long bitstreams typically require multiple cycles for a single multiplication. 

Hybrid stochastic-binary designs have been proposed to combine the energy efficiency of SC with the accuracy of binary computing~\cite{date17-hsb,isocc21,tvlsi-HSN}.
Bent-Pyramid (BP) data format is a quasi-stochastic data representation which originally proposed by~\cite{BP}. 
The compressed BP data format (BP8) reduces the bit width from 10-bit to 8-bit by removing redundant bits~\cite{OISMA,DISCA}.
Although BP8 is deterministic code without SNGs, it can still use the simple AND gate to perform SC multiplication. For each multiplication, BP multiplicands and multipliers are mapped differently to two complementary pre-designed datasets to keep a minimum correlation, thereby maintaining the multiplication accuracy. 

We propose leveraging this complementary trade-off to develop a hybrid design that combines BC and SC, thereby achieving a higher energy efficiency and classification accuracy. Our proposed architecture, BenDi, is a quasi-stochastic systolic architecture that eliminates the SNGs overhead by using signed BP (BP9) data format while maintaining low latency and high energy efficiency through DiP systolic dataflow ~\cite{dip} for matrix multiplication. Fig.~\ref{fig:f1} shows the abstract overview of our hardware-software co-design workflow.
This paper proposes the following contributions:
\begin{enumerate}
    \item A hardware-aware quantization of CNNs tailored for the BP9 data format, including Post-Training Quantization (PTQ) and Quantization-Aware Training (QAT).
    \item A quasi-stochastic systolic architecture that integrates BP9 and DiP dataflow, while leveraging customized tiling and data mapping techniques to handle the CNN's workloads.
    \item A hardware implementation of the architecture using commercial 22nm technology node at a low supply voltage with post-place-and-route evaluation of area, power consumption, and energy; comparing against the SOTA hardware accelerators.
\end{enumerate}

The rest of this paper is organized as follows. Section~\ref{sec:design} presents the proposed BenDi architecture, Section~\ref{sec:eval} reports the implementation results, and Section~\ref{sec:concl} concludes the paper.

\section{BenDi Architecture}
\label{sec:design}
\subsection{Hardware-Aware Quantization For BP9}
Given that BP8 is an unsigned data format and is not optimized for end-to-end model inference, we propose a signed version, BP9, by adding an extra sign bit. 
BP9 cannot be directly applied to existing binary quantization frameworks.
A custom Pytorch kernel was implemented for BP9 matrix multiplication and BP9 convolution to mimic the hardware operations, targeting a light 1D-CNN model. The traditional activation function was replaced by the $HardTanh$ function, which is more compatible with the data range of BP9.
During PTQ, we replace the native function with our custom kernel to perform inference with BP9.
To enable QAT for BP9, we adopt a straight-through estimator (STE)-based design. In the forward pass, weights/activations are quantized to the nearest value in the BP9 domain, and the matrix multiplication is performed using the BP9 custom kernel. In the backward pass, identity and standard matrix multiplication gradients are used to approximate the non-differentiable quantization and hardware.

\begin{figure}[!t]
    \centering
    \includegraphics[width=1\linewidth]{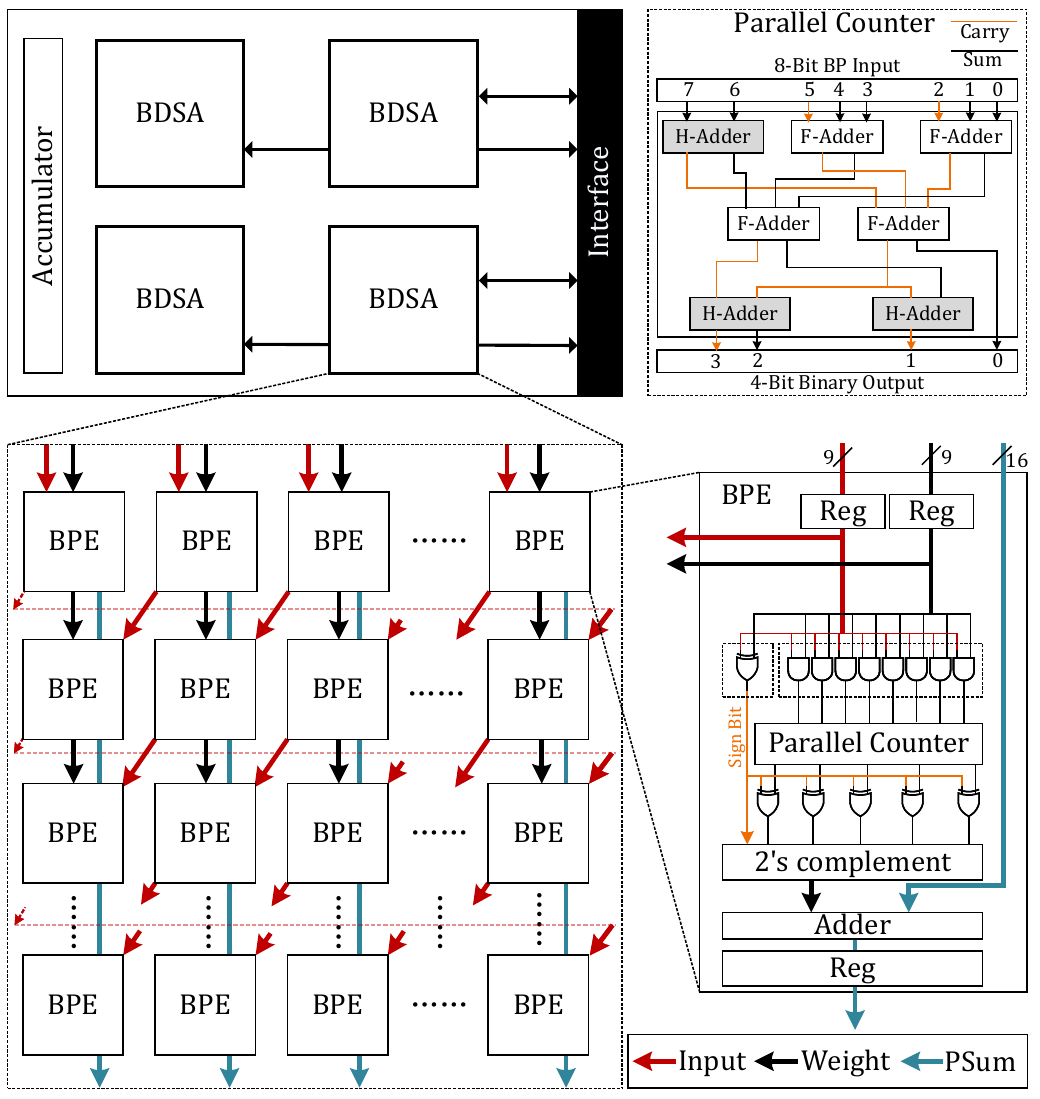}
    \caption{Overview of BenDi Systolic Architecture. It consists of 4 BDSAs with an accumulator. The zoom-in detail shows DiP-dataflow in systolic array, BPE architecture and PC architecture.}
    \label{fig:arch}
\end{figure}

\subsection{BenDi Systolic Architecture}

Fig.~\ref{fig:arch} shows the proposed BenDi hardware architecture, which incorporates 4 BenDi systolic arrays (BDSA), an accumulator. Each BDSA consists of 16x16 BP9 processing elements (BPEs). The BDSA performs multiplications in the SC domain and accumulation in the binary domain. By employing customized tiling and data mapping techniques, multiple BDSAs can operate in parallel, thereby increasing the hardware utilization and reducing the overall latency.
Every BDSA supports DiP dataflow, which is a SOTA weight stationary dataflow for binary-based matrix multiplication~\cite{dip}. DiP uses diagonal input and permutated weights to eliminate input and output synchronization buffers, thereby improving significantly the energy and area efficiencies. However, it is worth noting that the proposed BPEs are fundamentally orthogonal to the systolic dataflows, and can be integrated into any systolic dataflow.

To support signed BP data format, each BPE support 9-bit BP activation input and 9-bit BP weight input which 1 bit for sign and the other 8 bits for the BP8 format. The weights are pre-loaded down to the array directly, while input activations are diagonally propagated row by row. Instead of using conventional binary-based multipliers, BPE uses 8 AND gates to perform the SC multiplication, and an XOR gate to handle the sign bit. Then, a parallel counter (PC), composed of full adders (F-Adder) and half adders (H-Adder), counts the number of 1s and converts the SC multiplication results into the binary domain, as shown in Fig.~\ref{fig:arch} top right. The output sign bit and 5 XOR gates will handle the 2's complement conversion for a negative PC output, and then be accumulated by a conventional adder with a partial sum from the BPE in the same column and previous row.

\subsection{CNN Mapping on BenDi}
\begin{figure}[!t]
    \centering
    \includegraphics[width=1\linewidth]{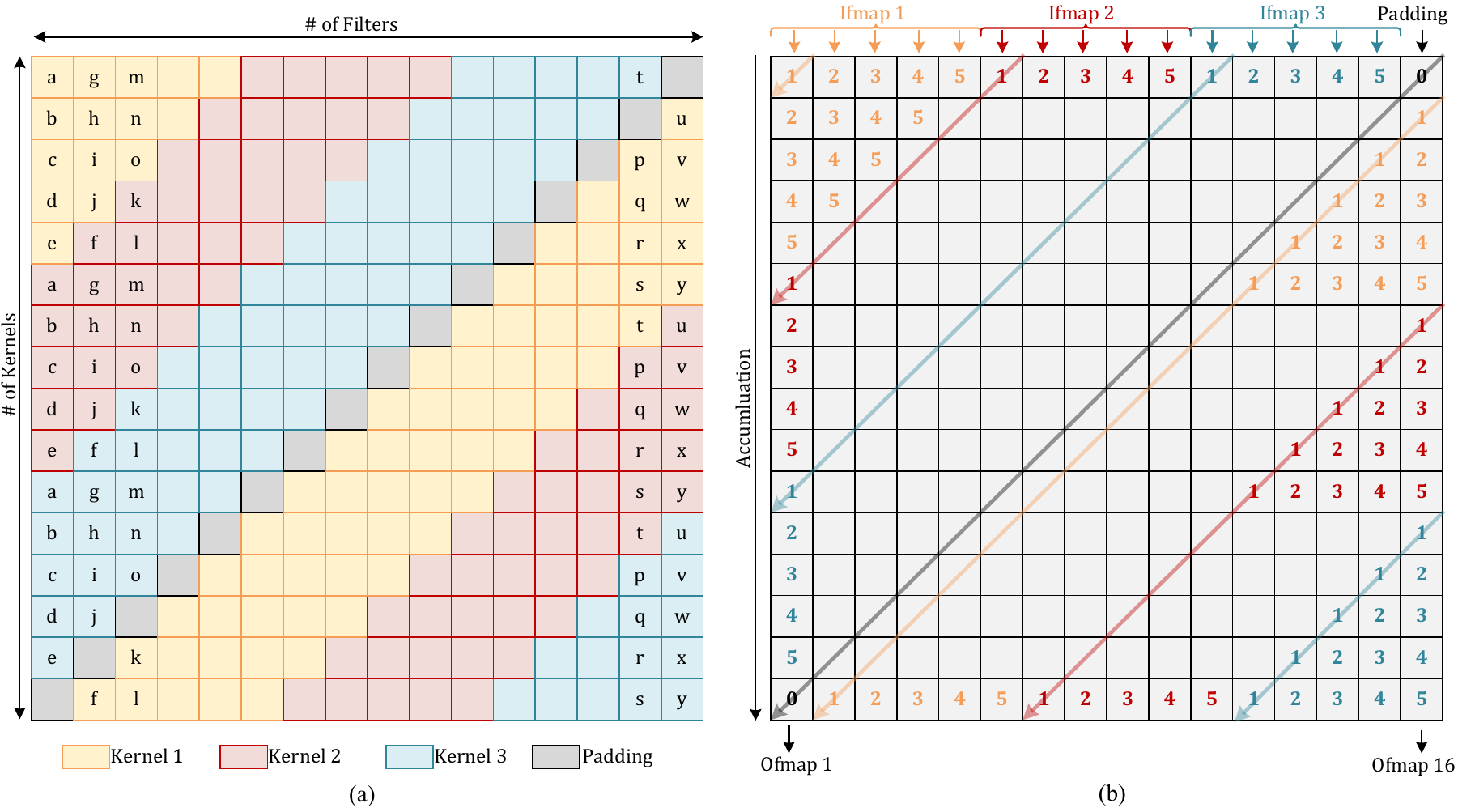}
    \caption{Mapping strategy for Conv layer on BDSA. (a) For weights, kernels within the same filter are arranged in a single column and permutated by row numbers vertically. (b) For activation mapping, the activations are diagonally distributed across the array so that each column produces one output feature map.}
    \label{fig:map}
\end{figure}

DiP was originally proposed as a square systolic array for matrix multiplication, not convolutional workloads~\cite{dip}. Our data mapping strategies for the convolutional layers (Conv) of CNNs are shown in Fig.~\ref{fig:map}. The Conv layer first adopts the im2col algorithm to convert the convolution into a matrix-multiplication operation. 
The weights mapping on BDSA is shown in Fig.~\ref{fig:map}(a), where each column corresponds to different kernels (different colors) within the same filter (same letters), to contribute to the same output feature map (Ofmap). 
Input activations (numbers) of different input feature maps (Ifmap) (different colors) are propagated diagonally through the BPEs, allowing different columns to handle different Ofmap channels and different filters, as shown by Fig.~\ref{fig:map}(b). In this manner, BDSA enables direct sharing of the different Ifmap activations among the different filters and kernels within the same systolic array, thereby maximizing not only the hardware utilization but also the data utilization locally within the array. 
     
\section{Evaluation and results}
\label{sec:eval}
To evaluate BenDi's performance, we train three CNN models for: (1) MNIST dataset, (2) MIT-BIH arrhythmia dataset~\cite{mitbih}, and (3) PhysioNet Apnea-ECG dataset~\cite{apnea}. After pre-processing, the MIT-BIH dataset has 109446 samples with 5 classes, and PhysioNet Apnea-ECG has 33654 samples with 2 classes. 
For the hardware evaluation, the architecture was implemented using SystemVerilog and the RTL design was synthesized with Cadence Genus at 100 MHz. We performed place-and-route with Cadence Innovus to estimate area and power consumption at 0.5 V supply voltage in a commercial 22nm technology node.

\subsection{Software Accuracy}
To investigate the accuracy, we trained a LeNet-like architecture for MNIST and two 1D-CNN models for MIT-BIH and Apnea-ECG datasets. For MIT-BIH, Conv1 and Conv2 use 16 and 32 filters, respectively, while for Apnea-ECG, they use 32 and 64 filters, respectively. Then, the 32-bit floating-point model was quantized to a BP9 model with our custom PTQ and QAT methods. Also, a native INT8 quantization model was used as a baseline for further comparison.

In Table~\ref{tab:acc_comparison}, we use accuracy as metrics for all datasets and the F1 score specially for the two biomedical datasets. The F1 score provides a balanced evaluation of a model in imbalanced classification tasks, especially in biomedical datasets. In direct PTQ, BP9 is marginally comparable to INT8 for all benchmarks. 
Compared to FP32, the PTQ-BP9 shows only 1.18\% accuracy loss in the MNIST dataset, but has 10.45\% and 6.32\% accuracy loss in the Apnea-ECG and MIT-BIH datasets, respectively. Thanks to the QAT-BP9 approach, the accuracy is increased by 7.17\% for the Apnea-ECG and by 5.23\% for the MIT-BIH datasets. Meanwhile, the F1 score is also increased by 0.058 for the Apnea-ECG and by 0.153 for the MIT-BIH datasets. These results show that BP9 outperforms INT8 for these benchmarks while slightly degrading the accuracy in comparison to the FP32 baseline by only 3.28\% for the Apnea-ECG and 1\% for the MIT-BIH. 

\begin{table}[!t]
\centering
\caption{Accuracy Comparison of Quantization over three datasets}
\label{tab:acc_comparison}
\begin{tabular}{lcccccc}
\toprule
\multirow{2}{*}{Methods} & MNIST &\multicolumn{2}{c}{Apnea-ECG} & \multicolumn{2}{c}{MIT-BIH}  \\
\cmidrule(lr){2-2}
\cmidrule(lr){3-4} 
\cmidrule(lr){5-6}
&ACC& F1 & ACC & F1 & ACC  \\
\midrule
FP32    & 98.84   & 0.811 & 87.59 & 0.90  & 97.89\\ 
\midrule
PTQ-INT8 & 98.63  & 0.754 & 77.28 & 0.824 & 96.57 \\
\midrule
PTQ-BP9  & 97.45 & 0.771 & 77.14 & 0.701&91.66\\
\midrule
\textbf{QAT-BP9} & - & \textbf{0.829} & \textbf{84.31} & \textbf{0.854} & \textbf{96.89} \\ 
\bottomrule
\end{tabular}
\end{table}

\subsection{Latency and Energy Results}

\begin{table}[!t]
\centering
\begin{threeparttable}
\caption{The Performance of BenDi for MIT-BIH and Apnea-ECG benchmarks, at 0.5V and 100MHz.}
\label{tab:energy}
\begin{tabular}{lcccccc}
\toprule
\multirow{2}{*}{}  &\multicolumn{3}{c}{MIT-BIH} & \multicolumn{3}{c}{Apnea-ECG}  \\
\cmidrule(lr){2-4}
\cmidrule(lr){5-7}
&\#Arrays&Latency  & Energy &\#Arrays &Latency  & Energy  \\
\midrule
Conv1 & 1  &1.28  & 1.64  & 2 &4.8& 12.31\\
\midrule
Conv2 & 4 &3.36  &17.23  & 4  &9.6&49.24\\
\midrule
FC1  & 4 &4&20.51  &4 &7.84&40.21\\
\midrule
FC2  & 2 &0.48 &2.46  &2 &0.48 &1.23\\
\midrule
Total  &  &9.12 &41.84  & &22.72 &102.99\\
\bottomrule
\end{tabular}
\begin{tablenotes}
      \footnotesize
      \item Note: Latency is measured in $\mu$S. Energy is measured in nJ.
    \end{tablenotes}
  \end{threeparttable}
\end{table}

Table~\ref{tab:energy} shows the performance metrics of the BenDi architecture at 0.5V and 100 MHz for MIT-BIH and Apnea-ECG datasets. Regarding MIT-BIH as a showcase, BenDi utilizes only 1 BDSA for Conv1 and 2 BDSAs for FC2, due to the low dimensions of these workloads. For larger dimensions of Conv2 and FC1, BenDi utilizes the full hardware of the 4 BDSAs for computing. To maximize the hardware utilization, the FC2 latency is hidden by pre-loading the data of the Conv1 layer of the next batch, following a pipeline manner. This reduces the total latency and improves the overall energy efficiency.

\subsection{Hardware Comparison}
\begin{figure}[!t]
    \centering
    \includegraphics[width=1\linewidth]{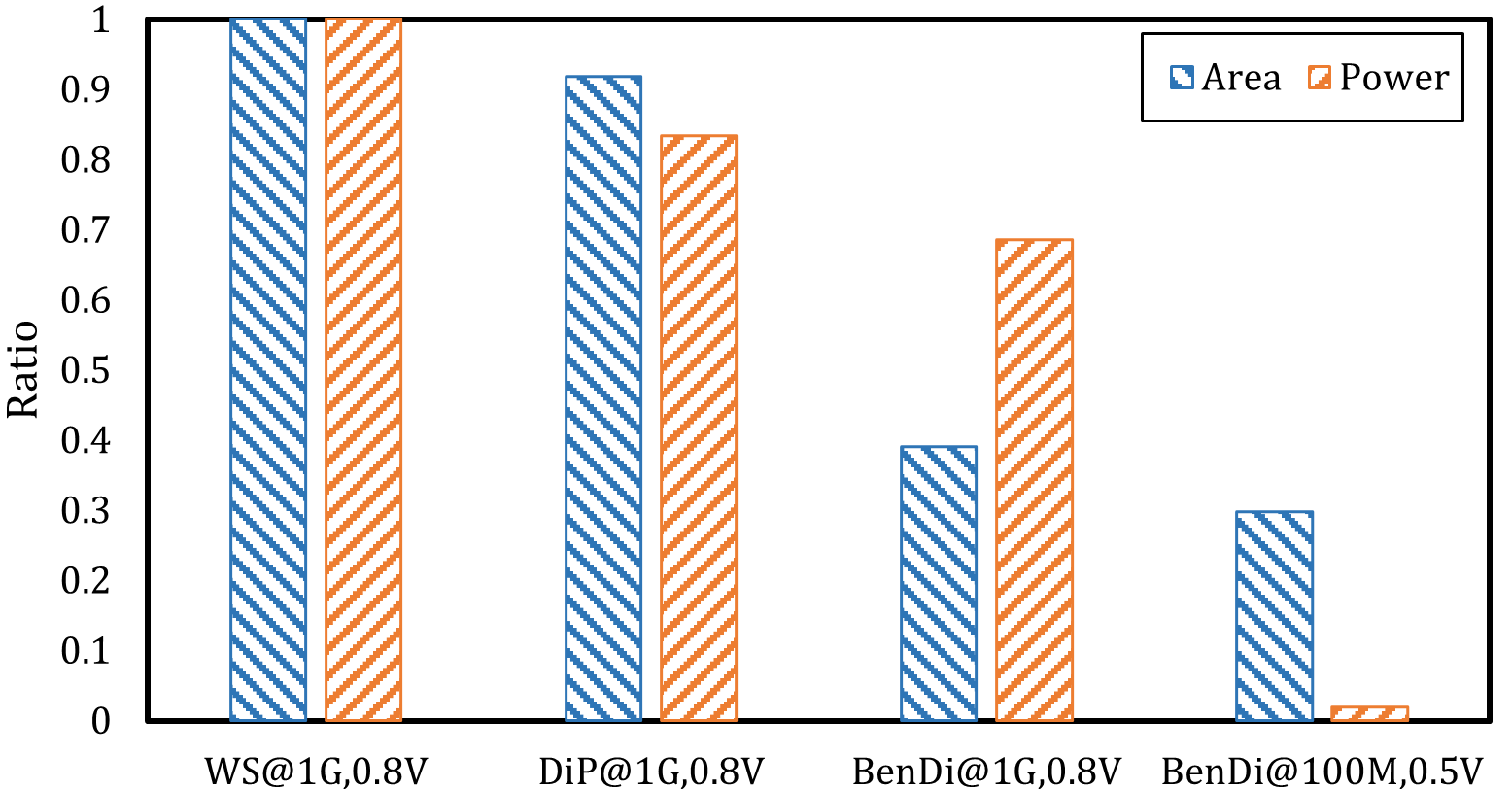}
    \caption{Normalized area and power results, comparing BenDi to WS and DiP systolic arrays. Results of WS and DiP are at 1 GHz and 0.8V, while BenDi's results are at 1 GHz/100 MHz and 0.8V/0.5V. All results are post-place-and-route of 16x16 systolic arrays using a commercial 22nm technology.}
    \label{fig:hw}
\end{figure}

The BenDi architecture was implemented, targeting two design points: (1) High Frequency, High Voltage for heavy workloads, and (2) Low Frequency, Low Voltage for edge applications. The two BenDi design points are compare against the conventional weight-stationary (WS) systolic array and the SOTA DiP implementation using the same technology node, as illustrated by Fig.~\ref{fig:hw}. At the same frequency (1 GHz) and supply voltage (0.8V), BenDi provides 2.35x/2.56x smaller area and 1.21x/1.45x lower power consumption, compared to DiP/WS, respectively. 
For a lower frequency (100 MHz) and lower voltage (0.5V), BenDi improves area by 3.0x/3.35x and power consumption by 41.8x/50.1x compared to DiP/WS, respectively, at 1GHz and 0.8V. 
Consequently, BenDi offers 4.1x and 5x better energy-efficiency compared to DiP and WS, respectively.

\begin{table}[!t]
\centering
 \begin{threeparttable}
\caption{Hardware Comparison with Other Bioelectronics Implementations.}
\label{tab:perf}
\begin{tabular}{lccc}
\toprule
  & \cite{tbcas1} &\cite{tbcas2} & \textbf{BenDi}\\
\midrule
Paradigm & Binary & Binary &  \textbf{Bent-Pyramid}  \\
\midrule
Model & 1D-CNN  & BNN  &  1D-CNN  \\
\midrule
Dataset & MIT-BIH & MIT-BIH  & MIT-BIH/Apnea-ECG  \\
\midrule
Data Width & 16-bit  & 1-bit  &  \textbf{9-bit}  \\
\midrule
Accuracy(\%) & 98.4  & 98.78  & \textbf{96.86}/84.31   \\
\midrule
Frequency (Hz) & 100M  & 100K  &  \textbf{100M}  \\
\midrule
Process (nm) & 55 & 55 &  \textbf{22}   \\
\midrule
Voltage (V) & 0.85 & 1.2& \textbf{0.5} \\
\midrule
Area (mm$^2$)  & 1.39  & 0.43& \textbf{0.085}  \\&(0.305*)&(0.204*)&\\
\midrule
Power (mW)  & 5.14  &0.00784 & \textbf{5.1299} \\&(3.212*)&(0.00482*)&\\
\midrule
Throughput \\(GOPS)  & 2.04 & - & \textbf{204.8}\\
\midrule
Area Eff.  &  0.001 & - & \textbf{2.409}\\(TOPS/mm$^2$)&(0.099*)&&\\
\midrule
Energy Eff. &  0.397& -  & \textbf{39.92}\\(TOPS/W) &(0.635*)&&\\
\bottomrule
\end{tabular}
\begin{tablenotes}
      \footnotesize
      \item *: Technology scaling to 22nm according to DeepScaleTool~\cite{deepscale}.
    \end{tablenotes}
  \end{threeparttable}
\end{table}

Table~\ref{tab:perf} summarizes and compares performance metrics with other related works. All related works support the MIT-BIH dataset, but \cite{tbcas1} uses a 1D-CNN with 16-bit data width and \cite{tbcas2} uses a Binarized CNN with 1-bit data width. BenDi uses 1D-CNN with 9-bit BP data format, achieving only 1.54\%-1.92\% accuracy loss compared to other works in MIT-BIH. However, BenDi achieves significant improvement by 100x in throughput, improving the area efficiency by 24.33x, and energy efficiency by 62.86x, compared to \cite{tbcas1}. 

\section{Conclusion}
\label{sec:concl}
In this work, we present BenDi, a quasi-stochastic systolic architecture based on a signed Bent-Pyramid data format, utilizing the energy-efficient dataflow and a customized BP9 quantization approach for CNN, targeting bioelectronic systems on the edge. 
To minimize energy and power consumption, BenDi incorporates four levels of optimization. At the software level, a compact 1D-CNN model with a hardware-aware BP9 quantization technique are leveraged. At the dataflow level, BenDi utilizes DiP dataflow to eliminate synchronization buffering overhead typical in traditional systolic dataflows. BenDi runs a hybrid approach using the BP9 data format at the computing level. Finally, BenDi utilizes an advanced commercial 22nm technology node at  0.5V and 100 MHz to reduce the power consumption footprint at the circuit level. 
Comprehensive evaluations demonstrated that BenDi offers 3.35x lower area and 5x better energy-efficiency compared to traditional weight-stationary systolic arrays. 
Regarding biomedical benchmarks, BenDi achieves 24.33x higher area efficiency and 62.86x higher energy efficiency in comparison to its counterpart.


\bibliographystyle{IEEEtran}
\bibliography{rf1}

\end{document}